\documentclass[aps,twocolumn]{revtex4}
\usepackage{graphicx}
\usepackage{bm}
\usepackage{latexsym,amssymb,amsfonts,amsthm,amscd,amsmath}

\begin{document}
\title{Double Kelvin Wave Cascade in Superfluid Helium}
\author{G.~Boffetta$^{1}$, A.~Celani$^{2,1}$, D.~Dezzani$^{1}$, J.~Laurie$^3$, and S.~Nazarenko$^3$}
\affiliation{$^1$
Dipartimento di Fisica Generale and INFN,
Universit\`a degli Studi di Torino, v. Pietro Giuria 1, 10125, Torino, Italy \\
and CNR-ISAC, Sezione di Torino, c. Fiume 4, 10133 Torino, Italy \\
$^2$ CNRS, Institut Pasteur, Rue du docteur Roux 25, 75015 Paris, France\\
$^3$ Mathematics Institute, University of Warwick, Coventry CV4 7AL, UK }
\date{\today}
\begin{abstract}
We study the double cascade of energy and wave action in a local
model of superfluid vortex filaments. The model is obtained from
a truncated expansion of the 2D Local Induction Approximation
and it is shown to support six-wave interactions. We argue that,
because of the uncertainty in the vortex core profile, this model
has the same status of validity as the traditionally used Biot-Savart
model with cutoff, but it has advantage of being much simpler.
Our minimal model leads to a wave kinetic equation for which we predict existence of two
distinct power-law scaling in the spectrum, corresponding to a direct cascade
of energy and an inverse one of wave action.
Direct numerical simulations
confirm the theoretical predictions in the weak turbulence regime.
\end{abstract}

\maketitle

It is well known that a classical vortex filament
can support linear waves. These
were predicted by Kelvin more than one century ago and experimentally
observed about 50 years ago in superfluid $^{4}He$.
At very low temperature, where the friction induced by normal fluid
component can be neglected, Kelvin waves can be dissipated only at
very high frequencies by phonon emission \cite{V01}.
Therefore at lower frequency energy is transferred among different
wavenumbers by the nonlinear coupling.
This is the mechanism at the basis of the Kelvin wave
cascade which sustains superfluid turbulence \cite{S95,V00}.

In recent years, single vortex Kelvin wave cascade has attracted
much theoretical \cite{KS04}, numerical \cite{KVSB01,VTM03,KS05} and
experimental \cite{BLS06} attention.
Even within the classical one-dimensional vortex model, different
degrees of simplification are possible. For small amplitudes, the
vortex configuration can be described by a two component vector field, made of
the coordinates of the vortex line in the plane transverse to the
direction of the unperturbed filament. These depend on the single coordinate
that runs
along the filament.
This system of equations admits
a Hamiltonian formulation, dubbed the
two-dimensional Biot-Savart formulation (2D-BS), see (\ref{eq:h2d})
below.
Another, more drastic, simplification is obtained by considering local
interactions only. This leads to the local induction approximation
(LIA) which was originally derived starting from the full 3D-BS
\cite{AH65}.
The main limitation of LIA is that it generates an integrable system
with infinite conserved quantities, as it is equivalent to the nonlinear
Schr\"odinger equation \cite{Hasimoto72}, and therefore the resonant wave interactions are
absent  (at all orders) and one cannot
 reproduce the phenomenology of the
full system. For this reason LIA, despite its simplicity, is of
little help for the study of Kelvin wave turbulence.

In this paper we consider the simplest model for vortex
filament able to sustain a turbulent energy cascade. The model
is obtained in the limit of small amplitudes by a Taylor expansion
of the 2D-LIA. The truncation breaks the integrability of the Hamiltonian
and therefore generates a dynamical system with two inviscid
invariants (energy and wave action). For this class of systems,
whose prototype is the two-dimensional Navier-Stokes
turbulence \cite{KM80}, we expect a dual cascade phenomenology
in which one quantity flows to small scales generating a
{\it direct cascade} while the other goes to larger scales
producing an {\it inverse cascade}. The possibility of a dual cascade
scenario for Kelvin waves turbulence has been recently suggested
\cite{lebedev,N06} but never observed.
Direct numerical simulations at high resolution confirm the dual
cascade picture with spectral exponents consistent with the
theoretical prediction based on a six-waves kinetic equation.

At a macroscopic level, the superfluid vortex filament is a classical
object whose dynamics is often described by the Biot-Savart equation (BSE)
\begin{equation}
\dot{\bf r} = {\kappa \over 4 \pi} \int {d {\bf s} \times ({\bf r}-{\bf s})
\over |{\bf r}-{\bf s}|^3}
\label{eq:bse}
\end{equation}
which describes the self-interaction of vortex elements.
The quantum nature of the phenomenon
is encoded in the discreteness of circulation $\kappa=h/m$ \cite{Donnelly91}.

The BSE dynamics of the vortex filament admits a Hamiltonian formulation
under a simple geometrical constraint: the position ${\bf r}$ of
the vortex is represented in two-dimensional parametric form
as ${\bf r}=(x(z),y(z),z)$, where $z$ is a given axis.
From a geometrical point of view, this corresponds to small
perturbations with respect to the straight line configuration, i.e.
the vortex cannot form folds in order to preserve the single-valuedness
of the $x$ and $y$ functions.
In terms of the complex canonical coordinate $w(z,t)=x(z,t)+i y(z,t)$,
BSE can be written in Hamiltonian form $i \dot{w} = \delta H[w]/\delta w^*$
with \cite{S95}
\begin{equation}
H[w] = {\kappa \over 4 \pi} \int dz_1 dz_2 {1 + Re(w'^{*}(z_1) w'(z_2))
\over \sqrt{(z_1 - z_2)^2 + |w(z_1)-w(z_2)|^2}}
\label{eq:h2d}
\end{equation}
where we have used the notation $w'(z)=\partial w/\partial z$.
The geometrical constraint of small amplitude
perturbation can be expressed in terms of a parameter
$\epsilon(z_1,z_2)=|w(z_1)-w(z_2)|/|z_1 - z_2| \ll 1$

An enormous simplification, both for theoretical and numerical
purposes, is obtained by means of the so called local induction approximation
(LIA) \cite{AH65}. This approximation is justified by the observation
that (\ref{eq:bse}) is divergent as ${\bf s} \to {\bf r}$ and
is obtained by introducing a cutoff at $a=|{\bf r}-{\bf s}|$
in the integral in (\ref{eq:bse}) which represents
the vortex filament radius. However,
the cutoff operation does not take into account 
the distribution of the vorticity inside the core of the
vortex filament. Indeed, let us consider linear Kelvin waves on an infinite
straight filament with wavevector $k$ such that $ka \ll 1$.
Dispersion relation for the frequency of such
linear waves can be found exactly from the Euler equations for several
special types of the vortex core shapes:
\begin{equation}
\omega_k=\frac{\kappa k^2}{4\pi}\left[\ln\left(\frac{1}{ka}\right)+C\right],
\label{eq:dispersion}
\end{equation}
where $C= -\gamma+\ln 2 +\frac{1}{4}$ for a vortex with uniform vorticity
inside a cylinder of radius $a$ \cite{Schwarz85} (where $\gamma$ is the Euler
constant),
$C= -\gamma + \ln 2 $  for a hollow core \cite{Barenghi06}, and
$C= -\gamma -\frac{3}{2}$ for the BSE model (\ref{eq:bse}) with cutoff
(see Appendix A).
From (\ref{eq:dispersion}) we see that the particular shape of the vortex
core affects the immediate next order with respect to the LIA term (log)
and, therefore, it affects the leading order of the {\em nonlinear}
dynamics. 
On the other hand,  the choice of the vortex core shape is
difficult and uncertain, considering that the fluid description itself 
fails within the vortex core. Thus, 
in the following we will use the
simplest model which arises naturally from the truncated expansion
of the LIA Hamiltonian.

When applied to Hamiltonian (\ref{eq:h2d}), the LIA procedure gives
\begin{equation}
H[w] = 2 {\kappa \over 4 \pi} \ln\left({\ell \over \xi}\right)
\int dz \sqrt{1 + |w'(z)|^2}
= 2 \beta L[w]
\label{eq:hlia}
\end{equation}
where $\ell$ is a length of the order of the
curvature radius (or intervortex distance when the considered vortex filament is a
part a vortex tangle),
$\beta=(\kappa/4 \pi)\ln(\ell/a)$. Here, it was taken into account that because
$a$ is much smaller than any other characteristic size in the system,
$\beta$ will be about the same whatever characteristic scale $\ell$ we take in its definition.
We remark that in the LIA approximation the Hamiltonian is proportional
to the vortex length $L[w]=\int dz \sqrt{1 + |w'(z)|^2}$ which is
therefore a conserved quantity.
The equation of motion from (\ref{eq:hlia}) is (we set $\beta=1/2$
without loss of generality, i.e. we rescale
time as $2 \beta t \to t$)
\begin{equation}
\dot{w} = {i \over 2} \left( {w' \over \sqrt{1+|w'|^2}} \right)'
\label{eq:lia}
\end{equation}
As a consequence of the invariance under phase transformations,
equation (\ref{eq:lia}) conserves also the total wave action
(also called the kelvon number) \cite{KS04}
\begin{equation}
N[w] = \int dz |w|^2
\label{eq:n}
\end{equation}

In addition to these two conserved quantities, the 2D-LIA model
possesses an infinite set of invariants and is
integrable, as it is the LIA of BSE (which can be transformed into
the nonlinear Schr\"odinger equation by the Hasimoto transformation, see Appendix B).
Integrability is broken if one considers a truncated expansion of
the Hamiltonian (\ref{eq:hlia}) in power of wave amplitude $w'(z)$.
Taking into account the lower order terms only one obtains:
\begin{eqnarray}
H_{exp}[w] &=& H_0 + H_1 + H_2 =  \nonumber \\
&=& \int dz \left( 1 + {1 \over 2} |w'|^2 - {1 \over 8} |w'|^4 \right)
\label{eq:expansion}
\end{eqnarray}
Neglecting the constant term, the Hamiltonian can be written in
Fourier space as
\begin{eqnarray}
H_{exp}&=&\int \omega_k |w_k|^2 dk + \nonumber \\
&+& \int dk_{1234} W_{1234} \delta^{12}_{34}(k) w_1^* w_2^* w_3 w_4
\label{eq:expfourier}
\end{eqnarray}
with $\omega=k^2/2$, $W_{1234}=-\frac{1}{8} k_1 k_2 k_3 k_4$ and
we used the standard notation
$\delta^{12}_{34}(k)=\delta(k_1+k_2-k_3-k_4)$ and
$dk_{1234}=dk_1 dk_2 dk_3 dk_4$.
For the dispersion relation $\omega \sim k^2$
there is no non-trivial solution to the four-wave resonant
conditions. There is no five-wave interaction either, because there are
no odd-degree Hamiltonians. The first non-vanishing process appears
to be six-wave.
Introducing a near-identity (weakly nonlinear) canonical transformation $w_k \to c_k$
using the standard strategy described in \cite{ZLF92}, it is possible to
transform (\ref{eq:expfourier}) into (See Appendix C)
\begin{eqnarray}
& & H_c=\int \omega_k |c_k|^2 dk + \nonumber \\
&+& \int dk_{123456} C_{123456}
\delta^{123}_{456}(k) c_1^* c_2^* c_3^* c_4 c_5 c_6
\label{eq:hc}
\end{eqnarray}
where $C_{123456}= -\frac{1}{16} k_1 k_2 k_3 k_4 k_5 k_6$. This six-order interaction coefficient
 is obtained from coupling of two fourth-order vertices of  $H_2$.
It is not surprising that the resulting expression coincides, with the opposite sign,
the interaction coefficient of $H_3$ in
(\ref{eq:lia}). Indeed, (\ref{eq:lia}) is an integrable model which implies that
if we retained the next order too, i.e. $H_3$, then the resulting six-wave process would
be nil, and the leading order would be an 8-wave process in this case.
Thus, the existence of the six-wave process is a
consequence of the truncation (\ref{eq:expansion}) of the Hamiltonian.

Hamiltonian (\ref{eq:expfourier}) (or equivalently (\ref{eq:hc})) constitutes the
minimal model for Kelvin wave turbulence.
It possesses the same scaling properties as the BSE system: it conserves the
energy and the wave action, it gives rise to a six-wave system with an interaction
coefficient with the same order of homogeneity as the one of the BSE.
A slight further modification should be made in the time re-scaling factor as
$\beta=(\kappa/4 \pi)$, - i.e. by dropping the large log factor from the original
definition (to have  the correct with respect to this log in rate of the six-wave process).
This minimal model is much simpler than BSE and it has the same degree of validity
as BSE because of the vortex core uncertainty discussed above.

Physical insight on Kelvin wave turbulence is obtained
from the wave turbulence (WT) approach which yields
 the kinetic equation which describes the dynamics of
the wave action density $n_k=\langle |c_k|^2 \rangle$.

The dynamical equation for the variable $c_k$ can be derived from Hamiltonian (\ref{eq:hc}) by the relation $i\partial c_k / \partial t = \delta H_c / \delta c^*_k$, and is
\begin{equation}\label{eq:cdot}
	i\frac{\partial c_k}{\partial t} - \omega_k c_k = \int dk_{23456}C_{k23456}c_2c_3c_4c^*_5c^*_6\delta^{k23}_{456}(k).
\end{equation}

Multiplying equation (\ref{eq:cdot}) by $c_k^*$, subtracting the complex conjugate and averaging we arrive at
\begin{equation}\label{eq:n_k}
\frac{\partial \langle c_kc_k^*\rangle}{\partial t}= 6Im\left(\int
dk_{23456} C_{k23456}J_{k23456}\delta^{k23}_{456}(k) \right),
\end{equation}
where $J_{k23456}\delta^{k23}_{456} = \langle c^*_kc_2^*c_3^*c_4c_5c_6 \rangle$.

Assuming a Gaussian wave field, one can take $J_{k23456}$ to the zeroth order $J^{(0)}_{k23456}$, which is simplified via the Gaussian statistics to a product of three pair correlators,
\begin{eqnarray}\label{eq:rpa}
J^{(0)}_{k23456}&=&n_2n_3n_4\big[\delta^k_4\left(\delta^2_5\delta^3_6+\delta^2_6\delta^3_5\right)\nonumber\\
&+&\delta^k_5\left(\delta^2_4\delta^3_6+\delta^2_6\delta^3_4\right)+\delta^k_6\left(\delta^2_4\delta^3_5+\delta^2_5\delta^3_4\right)\big].
\end{eqnarray}
However, due to the symmetry of $C_{k23456}$ this makes the right hand side of the kinetic equation (\ref{eq:ke}) zero.  To find a nontrivial answer we need to obtain a first
order addition $J^{(1)}_{k23456}$ to $J_{k23456}$. To calculate $J^{(1)}_{k23456}$ one takes the time derivative of $J_{k23456}$, using the equation of motion (\ref{eq:cdot}) and inserting the zeroth order approximation for the tenth correlation function (this is similar to equation \ref{eq:rpa}, but a product of five pair correlators involving ten wavevectors) $J^{(1)}_{k23456}$ can then be written as

\begin{equation}\label{eq:integratedJ}
J^{(1)}_{k23456}=Be^{i\Delta\omega t}+\frac{A_{k23456}}{\Delta\omega},
\end{equation}
where $\Delta\omega=\omega_k+\omega_2+\omega_3-\omega_4-\omega_5-\omega_6$ and $A_{k23456}= 3C^*_{k23456}n_k n_2 n_3 n_4 n_5 n_6 \left[{1 \over n_k}+{1 \over n_2}
+{1 \over n_3}-{1 \over n_4}-{1 \over n_5}-{1 \over n_6} \right]$.  The first term of (\ref{eq:integratedJ}) is a fast oscillating function, its contribution to the integral (\ref{eq:n_k}) decreases with $z$ and is negligible at $z$ larger than $1/\omega_k$, and as a result we will ignore the contribution arising from this term. The second term is substituted in equation (\ref{eq:n_k}),the relation $Im(\Delta\omega)\sim -\pi\delta(\Delta\omega)$ is applied because of integration around the pole, and the kinetic equation (\ref{eq:ke}) derived,

\begin{eqnarray}
\dot{n}_k &=& 18 \pi \int dk_{23456} \; |C_{k23456}|^2 \;
\delta^{k23}_{456}(k) \; \times \nonumber \\
&\times& \delta^{k23}_{456}(\omega) \; f_{k23456}
\label{eq:ke}
\end{eqnarray}
where we have introduced
$f_{k23456}=n_k n_2 n_3 n_4 n_5 n_6 \left[{1 \over n_k}+{1 \over n_2}
+{1 \over n_3}-{1 \over n_4}-{1 \over n_5}-{1 \over n_6} \right]$.

A simple dimensional analysis of (\ref{eq:ke}) gives
\begin{equation}
\dot{n}_k \sim k^{14} n_k^5
\label{eq:dim}
\end{equation}
which is the same form obtained from the full BSE \cite{KS04}.
The energy flux at wavenumber $k$ is defined as
$\Pi^{(H)}_k = \int dk' \dot{n}_{k'} \omega_{k'}$ which, using
(\ref{eq:dim}), becomes $\Pi^{(H)}_k \sim k^{17} n_{k}^5$.
By requiring the existence of a range of scales in which the energy flux is
$k$-independent leads to the spectrum
\begin{equation}
n_k \sim k^{-17/5}
\label{eq:spdir}
\end{equation}
A similar argument can be applied to the wave action (\ref{eq:n})
whose flux is $\Pi^{(N)}_k = \int dk' \dot{n}_{k'} \sim k^{15} n_{k}^5$.
Therefore a scale independent flux of wave action requires a spectrum
\begin{equation}
n_k \sim k^{-3}
\label{eq:spinv}
\end{equation}
The two spectra (\ref{eq:spdir}-\ref{eq:spinv}) occur in different
scale ranges and
the two cascades develops in opposite directions, as in the case
of two-dimensional turbulence \cite{KM80}. Among the two conserved
quantities, the largest contribution to energy comes from smaller scales
than those that contribute to wave action (because the former
contains the field derivatives). Therefore, according to
the Fj{\o}rtoft argument \cite{Fjortoft53}, we expect to have
a {\it direct cascade} of energy with $k^{-17/5}$ spectrum at large $k$ and
an {\it inverse cascade} of wave action with spectrum $k^{-3}$ at small $k$.

In the following we will consider numerical simulations of the
system (\ref{eq:expansion}) under the conditions in which
a stationary turbulent cascade develops. Energy and wave action are
injected in the vortex filament by a white-in-time external forcing
$\phi(z,t)$ acting on a narrow band of wavenumbers around a given $k_f$.
In order to have a stationary cascade, we need additional terms which
remove $H$ and $N$ at small and large scales. The equation of motion
obtained from (\ref{eq:expansion}) is therefore modified as
\begin{equation}
\dot{w} = {i \over 2} \left[ w' \left( 1 - {1 \over 2} |w'|^2 \right)
\right]' - (-1)^p \nu \nabla^{2p} w - \alpha w + \phi
\label{eq:liatot}
\end{equation}
In (\ref{eq:liatot})
the small scale dissipative term (with $p > 1$) physically
represents the radiation of phonons (at a rate proportional to
$\nu$) and the large scale damping term can be interpreted
as the friction induced by normal fluid at a rate $\alpha$.

Assuming the spectra (\ref{eq:spdir}-\ref{eq:spinv})
a simple dimensional analysis gives the IR and UV cutoff induced
by the dissipative terms. The direct cascade is removed
at a scale $k_{\nu} \sim \nu^{-5/(10p-2)}$ while the inverse cascade
is stopped at $k_{\alpha} \sim \alpha^{1/2}$. Therefore, in an
idealized realization of infinite resolution one would obtain a
double cascade by keeping $k_f=O(1)$ and letting $\nu,\alpha \to 0$.
In order to have an extended inertial range, in finite resolution
numerical simulations we will restrict to resolve a single cascade
by putting $k_f \simeq k_{\alpha}$ or $k_f \simeq k_{\nu}$ for
the direct and inverse cascade respectively.

We have developed a numerical code which integrates the
equation of motion (\ref{eq:liatot}) by means of pseudospectral
method for a periodic vortex filament of length $2 \pi$ with a
resolution of $M$ points.
Linear, dissipative terms are integrated explicitly
while the nonlinear term is solved by a second-order Runge-Kutta
time scheme. Vortex filament is initially a straight line
($w(z,t=0)=0$) and long time integration is performed until a
stationary regime (indicated by the values of $H$ and $N$)
is reached.
The ratio between the two terms in the expansion (\ref{eq:expansion})
is $H_1/H_2 \simeq 20$, confirming {\it a posteriori} the validity
of the perturbative expansion (\ref{eq:expansion}).

The first set of simulations is devoted to the study of the
direct cascade.
Energy fluctuations are injected at forcing wavenumber
$k_f \simeq 2$ and the friction coefficient $\alpha$ is set in order
to have $k_{\alpha} \simeq k_f$.
Energy is removed at small scales by hyperviscosity of order $p=4$ which
restricts the range of dissipation on a narrow range of wavenumber
close to $k_{max}$.

\begin{figure}[h]
\begin{center}
\includegraphics[scale=0.7]{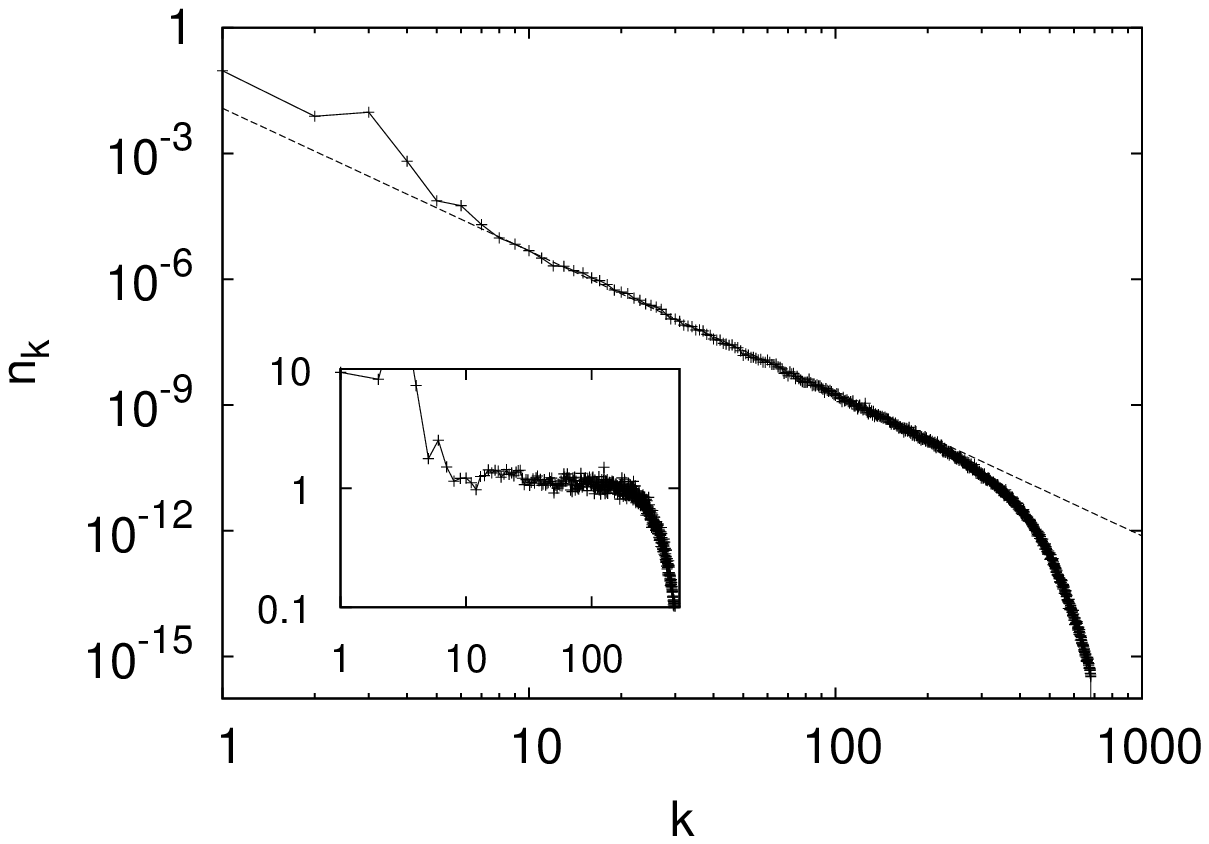}
\caption{Wave number spectrum $n_k$ for a simulation of the direct
cascade in stationary conditions at resolution $M=2048$. Forcing
is restricted to a range of wavenumbers $1 \le k_f \le 3$ and
phonon dissipation is modeled with hyperviscosity of order $p=4$.
The line represents the kinetic equation prediction
$n_k \simeq k^{-17/5}$. The inset shows the spectrum compensated
with the theoretical prediction.}
\label{fig2}
\end{center}
\end{figure}

In Figure~\ref{fig2} we plot the wave action spectrum for the
direct cascade run averaged over time in stationary conditions.
A well developed power law spectrum very close
to prediction (\ref{eq:spdir}) is observed over more than one
decade (see inset). This spectrum confirms the existence of
non trivial dynamics with six-wave process for the truncated
Hamiltonian (\ref{eq:expansion}).

We remind that the direct cascade for the Hamiltonian (\ref{eq:expansion})
was already discussed by Svistunov \cite{S95,KS05}
who also gave the dimensional prediction (\ref{eq:spdir}).
Numerical simulations with a discrete version of the Hamiltonian
(\ref{eq:expansion}), which breaks the integrability of the
original system as well, confirmed the validity of the dimensional
prediction \cite{KS05}.

\begin{figure}[h]
\begin{center}
\includegraphics[scale=0.7]{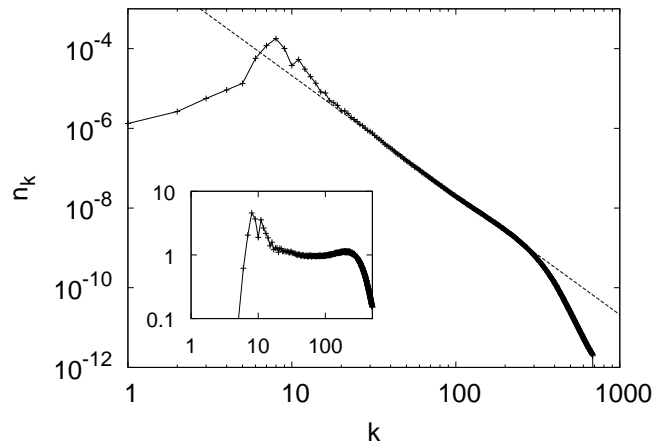}
\caption{Wavenumber spectrum $n_k$ for a simulation of the inverse
cascade in stationary conditions at resolution $M=2048$. Forcing
is restricted to a range of wavenumbers around $k_f=300$ and
phonon dissipation is modeled with hyperviscosity of order $p=4$.
The line represents the kinetic equation prediction
$n_k \simeq k^{-3}$. The inset shows the spectrum compensated
with the theoretical prediction.}
\label{fig3}
\end{center}
\end{figure}

We now turn to the simulations for the inverse cascade regime.
To obtain an inverse cascade, forcing is concentrated at small scales,
here $k_f=683$. In order to avoid finite size effects and accumulation
at the largest scale \cite{SY94}, the friction coefficient is chosen in such a
way that wave action is removed at a scale $k_{\alpha}\simeq 10$.
Figure~\ref{fig3} shows the spectrum for
this inverse cascade in stationary conditions.
In the compensated plot, a small deviation from the power-law scaling
at small scales, probably due to the presence of forcing, is observed.
Nevertheless, a clear scaling compatible with the dimensional analysis
of the kinetic equation is observed over about a decade.

In summary, we have introduced a minimal model for Kelvin wave turbulence,
and presented an argument why this model should be preferential over BSE.
Namely, this model is much simpler than BSE and it has the same
degree of validity due to the vortex core shape sensitivity and uncertainty.
We have used our model for numerical simulations of the direct and the
inverse cascades and found spectra which are in very good agreement with the
predictions of the WT theory.

\section{Appendix A - Interaction Coefficients of 2D-BS}

In this Appendix we extend the work of Kozik and Svistunov on the Kelvin wave cascade (KS$04$) \cite{KS04}.  They consider the full Biot-Savart Hamiltonian (\ref{eq:h2d}) in $2$D, and simplify the denominator by Taylor expansion.  The criterion for Kelvin-wave turbulence is that wave amplitude is small compared to wavelength, this is formulated as:
\begin{equation}
\epsilon(z_1,z_2)=\frac{|w(z_1)-w(z_2)|}{|z_1-z_2|}\ll 1.
\end{equation}

KS$04$ find the Biot-Savart Hamiltonian (\ref{eq:h2d}) expanded in powers $\epsilon$ ($H={H}_0+{H}_1+{H}_2+{H}_3$, here ${H}_0$ is just a number and is ignored) is represented as:

\begin{align}\label{eq:KS04Ham}
{H}_1&=&\frac{\kappa}{8\pi}\int \frac{d z_1 d z_2}{|z_1-z_2|}\left[2Re\left(w^{'*}(z_1)w^{'}(z_2)\right)-\epsilon^2\right],\nonumber\\
{H}_2&=&\frac{\kappa}{32\pi}\int \frac{d z_1 d z_2}{|z_1-z_2|}\left[3\epsilon^4-4\epsilon^2Re\left(w^{'*}(z_1)w^{'}(z_2)\right)\right],\\
{H}_3&=&\frac{\kappa}{64\pi}\int \frac{d z_1 d z_2}{|z_1-z_2|}\left[6\epsilon^4Re\left(w^{'*}(z_1)w^{'}(z_2)\right)-5\epsilon^6\right]\nonumber.
\end{align}

By taking the Fourier transformation $w(z)=\int dk w_k e^{ikz}$ of (\ref{eq:KS04Ham}), invoking a cutoff at $|z_1-z_2|=a$ because of the singularity present in the Biot-Savart Hamiltonian (\ref{eq:h2d}) as $|z_1-z_2|\rightarrow 0$, KS$04$ derived the coefficients of $H_1$, $H_2$ and $H_3$ in terms of cosines in Fourier space \cite{KS04}.  Kozik and Svistunov however, left their work unfinished.  We extend the work of KS$04$ by integrating the Fourier representations of $H_1$, $H_2$ and $H_3$ using integration by parts, and applying the following cosine identity \cite{GR80}

\begin{eqnarray}
\int_a^\infty\frac{\cos(t)}{t}dt &=& -\gamma-\ln(a)-\int_0^a\frac{\cos(t)}{t}dt\nonumber\\
&=& -\gamma-\ln(a)-\sum_{k=1}^\infty\frac{\left(-a^2\right)^k}{2k\left(2k\right)!}\nonumber\\
&=& -\gamma-\ln(a)+\mathcal{O}(a^2)
\end{eqnarray}
Neglecting terms of order $\sim a^2$ and higher we calculate the frequency and interaction coefficients of equations (\ref{eq:KS04Ham})
\begin{eqnarray}
{H}_1 &=&\int \omega_k |w_k|^2 d k,\\
{H}_2 &=&\int d k_{1234}W_{1234}w_1w_2w^*_3w^*_4\delta^{12}_{34}(k) ,\\
{H}_3 &=&\int d k_{123456}C_{123456}w_1w_2w_3w^*_4w^*_5w^*_6\delta^{123}_{456}(k) .
\end{eqnarray}
where
\begin{eqnarray}
\omega_k &=& \frac{\kappa}{4\pi}k^2\left[\ln\left(\frac{1}{ka}\right)-\gamma -\frac{3}{2}\right],\label{eq:BSomega}\\
W_{1234} &=& \frac{\kappa}{64\pi}k_1k_2k_3k_4\left[1+4\gamma-4\ln\left(\frac{1}{k_{\mathrm{eff}}a}\right)\right]\nonumber\\
 &+& F_{1234},\label{eq:BSW}\\
C_{123456}&=& \frac{\kappa}{128\pi}k_1k_2k_3k_4k_5k_6\left[1-4\gamma+4\ln\left(\frac{1}{k_{\mathrm{eff}}a}\right)\right]\nonumber\\
 &+& G_{123456}.\label{eq:BST}
\end{eqnarray}
We use the notation that $k_{\mathrm{eff}}$ is the mean value of wavenumbers, $\gamma=0.5772\dots$ is the Euler Constant and $F_{1234}$ and $G_{123456}$ are logarithmic terms of order one shown below
\begin{eqnarray}
F_{1234}&=&\frac{\kappa}{16\pi}\Big[6\sum_{N \in I}\frac{N^4}{24}\ln\left(\frac{N}{k_{\mathrm{eff}}}\right)\nonumber\\
 &-& \sum_{N \in J}k_ik_j\frac{N^2}{2}\ln\left(\frac{N}{k_{\mathrm{eff}}}\right)\Big]
\end{eqnarray}
\begin{eqnarray}
G_{123456}&=&\frac{\kappa}{16\pi}\Big[3\sum_{N \in K}k_6k_2\frac{N^4}{24}\ln\left(\frac{N}{k_{\mathrm{eff}}}\right) \nonumber\\
&-& 5\sum_{N \in L}\frac{N^6}{720}\ln\left(\frac{N}{k_{\mathrm{eff}}}\right)\Big]
\end{eqnarray}
\begin{align}
I=&\left\{ -[_1],-[_2],-[^3],-[^4],[^3_2],[^{43}],[^4_2] \right\}\\
J=&\Big\{ \left\{[^4], [_1], -[^{43}],-[^4_2]\right\}_{i=4,j=1},\nonumber\\
&\left\{[^3], [_1], -[^{43}],-[^3_2]\right\}_{i=3,j=1},\nonumber\\
&\left\{[^3], [_2], -[^{43}],-[^3_1]\right\}_{i=3,j=2},\nonumber\\
&\left\{[^4], [_2], -[^{43}],-[^4_1]\right\}_{i=4,j=2}\Big\}\\
K=&\Big\{ [_2],-[^5_2],-[_{23}],[^5_{23}],-[^4_2],[^{45}_2],[^4_{23}],-[^6_1],[^6],-[^{56}],-[^6_3],\nonumber \\
&[^{56}_3],-[^{46}],[^{456}],[^{46}_3],-[_{12}]\Big\}\\
L=&\Big\{ -[^4],-[_1],[^4_1],-[^6],[^{46}],[^6_1],-[^{46}_1],-[^5],[^{45}],[^5_1],-[^{45}_1],\nonumber\\
&[^{65}],-[^{456}],-[^{56}_1],[_{23}],-[_3],[^4_3],[_{13}],-[^4_{13}],[^6_3],-[^{46}_3],\nonumber\\
&-[^6_{13}],[^5_2],[^5_3],-[^{45}_3],-[^5_{13}],[^6_2],-[^{65}_3],[_{12}],[^4_2],-[_2]\Big\}.
\end{align}
The notation used for the logarithmic terms is: for $N\in K = \pm[^{pq}_r]$ then the corresponding term in $G_{123456}$ is
\begin{displaymath}
\pm\frac{3\kappa}{16\pi}k_6k_2\frac{(k_p+k_q-k_r)^4}{24}\log\left(\frac{k_p+k_q-k_r}{k_{\mathrm{eff}}}\right).
\end{displaymath}
As you can see, the logarithmic terms $F_{1234}$ and $G_{123456}$ are of order $\sim 1$ and are extremely messy and complex.  Applying the parametrization of Zakharov and Schulman \cite{ZS82} does not help in this case (see Appendix C).  We know that LIA is fully integrable, i.e. that after applying the canonical transformation all terms of order $\ln(ka)$ will cancel.  However, the presence of the order one additions to the dispersion and interaction coefficients, will break this integrability and result in some six-wave interactions.  Due to us being unable to simplify the expression after the canonical transformation because of the non-canceling order one terms, we make the assumption of the interaction coefficients being products of wavenumbers, this is equivalent to the truncated LIA model suggested in this letter.

\section{Appendix B - Integrability of 2D-LIA}
We show the details of the derivation that the LIA of BSE yields the 2D-LIA model (\ref{eq:lia}), i.e. that the
cutoff operation commutes with making the 2D reduction.
In the view of integrability of the original LIA, this amounts in a proof of integrability of the
2D-LIA model (\ref{eq:lia}).

The LIA representation of BSE can be written as

\begin{equation}\label{eq:LIAofBSE}
\dot{\bf r}= \beta {\bf r}'\times {\bf r}''
\end{equation}
where the notation for the differentiation $'$ is $d/dl$ where $l =\left(1+|{\bf w}|^2\right)^{-1/2}$ is the arc length.
 The two-dimensional representation of a vortex line can be represented by a vector ${\bf r}=z\hat{z}+{\bf w}$.  The vector ${\bf w} = (x(z),y(z))$ being a function of $z$, orientated in the $xy$-plane.



Applying the chain rule to rewrite all derivatives to be with respect to $z$ (from this point on $'$ will refer to $d/dz$), LIA can be represented as
\begin{equation}\label{eq:rdot}
\dot{{\bf r}}=\beta\left[\left(1+|{\bf w}'|
^2\right)^{-3/2}(\hat{z}\times{\bf w}''+{\bf w}'\times{\bf w}'')\right].
\end{equation}
With a little geometrical intuition, one can show $\dot{{\bf w}}=\dot{\bf r} - (\dot{\bf r}\cdot\hat{z})(\hat{z}+{\bf w}')$ \cite{S95}. Both ${\bf w}'$ and ${\bf w}''$ are perpendicular to $\hat{z}$ direction, thus, one can represent ${\bf w}'\times {\bf w}''=(({ \bf w}'\times {\bf w}'')\cdot\hat{z})\hat{z}=A\hat{z}$. Then equation (\ref{eq:rdot}) reduces to
\begin{equation}\label{eq:wdotfinal}
\dot{\bf w}=\beta \left(1+|{\bf w}'|^2\right)^{-3/2}\left[\hat{z}\times{\bf w}''-A{\bf w}'\right].
\end{equation}

We will now show that equation (\ref{eq:lia}) is equivalent to equation (\ref{eq:wdotfinal}).  First we must change our representation of $w$ from a complex variable $w(z)=x(z)+iy(z)$ to vector notation ${\bf w}=(x(z),y(z))$.  Equation (\ref{eq:lia}) is equivalent to
\begin{equation}\label{eq:com2vec}
\dot{\bf w}=\frac{1}{2}\hat{z}\times \frac{\partial}{\partial z}\left(
\frac{{\bf w}'}{\sqrt{1+|{\bf w}'|^2}}\right).
\end{equation}
Expanding, keeping track of $\beta$ and applying the vector identity $({\bf a}\times{\bf b})\times {\bf c}=({\bf c}\cdot {\bf a}){\bf b}-( {\bf c}\cdot {\bf b}) {\bf a}$, equation (\ref{eq:com2vec}) can be re-written as
\begin{equation}
\dot{\bf w}=\beta \left(1+|{\bf w}'|^2\right)^{-3/2}\left[\hat{z}\times {\bf w}''-A{\bf w}'\right].
\end{equation}
This is exactly the same result as Equation (\ref{eq:wdotfinal}). The 2D-LIA model (\ref{eq:lia}) is equivalent to the LIA of BSE (\ref{eq:LIAofBSE}) and so the 2D-LIA model is indeed integrable.

\section{Appendix C - Canonical Transformation}
In Wave Turbulence, the near-identity transformation allows one
to eliminate ``unnecessary" lower orders of nonlinearity in the system
if corresponding order of the wave interaction process is nil \cite{ZLF92}.
For example, if there is no three-wave resonances, then one can eliminate
the cubic Hamiltonian.
(The quadratic Hamiltonian corresponding to the linear dynamics, of course, stays).
 This process can be repeated recursively, in a way similar to the KAM theory, until
 the lowest order of the non-trivial resonances is reached. If no such resonances appear
 in any order, one has an integrable system.

 In our case, there is no four-wave resonances (there are no non-trivial solution
 for the resonance conditions in for $\omega \sim k^x$ if $x>1$).
 There is also no five-wave resonances because the original terms in the Hamiltonian are
 of the even orders. However, there are nontrivial solutions of the six-wave resonant
 conditions. Thus, one can use the near-identity transformation to convert our system into
 the one with the lowest order interaction Hamiltonian to be of degree six. Let us do this.

 A trick for finding a shortcut derivation of such a transformation was found in
 \cite{ZLF92}. It relies on the fact that the time evolution operator is a canonical
 transformation.
   Taking the Taylor expansion of $w(k,t)$ around
$w(k,0)=c(k,0)$ we get a desired transformation, that is by its
derivation, canonical.  The coefficients of each term can be
calculated from an auxiliary Hamiltonian $H_{aux}$. Similar procedure was
done in Appendix A$3$ of \cite{ZLF92} to eliminate the cubic Hamiltonian in cases when the three-wave
interaction is nil, and here we apply a similar approach to eliminate the quadric Hamiltonian.
 The transformation is represented as
\begin{equation}\label{eq:trans}
w_k=c(k,0)+t\left(\frac{\partial c(k,t)}{\partial
t}\right)_{t=0}+\frac{t^2}{2}\left(\frac{\partial^2 c(k,t)}{\partial
t^2}\right)_{t=0}+\cdots
\end{equation}
The transformation is canonical for all $t$, so for simplicity we set $t=1$. The coefficients of (\ref{eq:trans}) can be calculated from the followinf formulae,

\begin{align}\label{eq:HamSys}
\left(\frac{\partial c(k,t)}{\partial
t}\right)_{t=0}&=-i\frac{\delta H_{aux}}{\delta c^*}\nonumber \\
 \left(\frac{\partial^2c(k,t)}{\partial t^2}\right)_{t=0}&=-i
\frac{\partial}{\partial t}\frac{\delta H_{aux}}{\delta c^*}.
\end{align}

Due to the original Hamiltonian (\ref{eq:expansion}) $H_{exp}$ having $U(1)$ symmetry we have no odd wave-interactions, this simplifies the canonical transformation (\ref{eq:trans}) greatly, because the absence of odd interaction in $H_{exp}$ automatically fixes the arbitrary odd interaction coefficients within the auxiliary Hamiltonian $H_{aux}$. Transformation (\ref{eq:trans}) reduces to

\begin{eqnarray}\label{eq:trans2}
w_k&&=c_k-
\frac{i}{2}\int d k_{234}\tilde{W}_{k234}c^*_2c_3c_4\delta^{k2}_{34}(k)\nonumber\\
&&-3i\int d
k_{23456}\tilde{C}_{k23456}\delta^{k23}_{456}(k)c^*_2c^*_3c_4c_5c_6\nonumber\\
&&+\frac{1}{8}\int d k_{234567}\Big(\tilde{W}_{k743}\tilde{W}^*_{7623}\delta^{k7}_{45}(k)\delta^{76}_{23}(k)\nonumber\\
&&-2\tilde{W}_{k247}\tilde{W}_{7356}\times\delta^{k2}_{47}(k)\delta^{37}_{56}(k)\Big)c^*_2c^*_3c_4c_5c_6 .
\end{eqnarray}
Here we have used tildes to represent interaction coefficients of the auxiliary Hamiltonian $H_{aux}$.  To eliminate the nonresonant interactions we substitute transformation (\ref{eq:trans2}) into Hamiltonian (\ref{eq:expfourier}), this will yield a new representation in variable $c_k$ for Hamiltonian (\ref{eq:expfourier}) where the nonresonant terms (more specifically the four-wave interaction terms) will involve both $W_{1234}$ and $\tilde{W}_{1234}$.  Arbitrariness of $\tilde{W}_{1234}$ implies that we can select this to eliminate the total four-wave nonresonant interaction term, $H_2$ present within the Hamiltonian.  In our case,
\begin{equation}
\tilde{W}^*_{1234}=\frac{4iW_{1234}}{\omega_1+\omega_2-\omega_3-\omega_4}.
\end{equation}
This choice is valid as the denominator will not vanish due to the nonresonance of four-wave interactions.  Hamiltonian (\ref{eq:expfourier}) expresses in variable $c_k$, $H_{c}$ becomes

\begin{eqnarray}\label{eq:fullham}
H_{c}&&=\int \omega_k c_kc_k^* d
k\nonumber\\
&&+\int
\Big(C_{123456}-i\left(\omega_1+\omega_2+\omega_3-\omega_4-\omega_5-\omega_6\right)\nonumber\\
&&\times\tilde{C}_{123456}\Big)\delta^{123}_{456}(k)c^*_1c^*_2c^*_3c_4c_5c_6 d
k_{123456},
\end{eqnarray}

$\tilde{C}_{123456}$ is the arbitrary six-wave interaction coefficient from the auxiliary Hamiltonian.  This term does not contribute to the six-wave resonant dynamics as the factor in front will vanish on the resonant manifold that appears in the kinetic equation.  However, we can select $i\left(\omega_1+\omega_2+\omega_3-\omega_4-\omega_5-\omega_6\right)\tilde{C}_{123456}$ to equal $C_{123456}$ off the resonant manifold.  This enables us to write $H_{c}$ as equation (\ref{eq:hc}), where the explicit formula for $C_{123456}$ stemming from the transformation is
\begin{equation}\label{eq:C123456}
\begin{array}{rl}
C_{123456}=\frac{1}{18}\displaystyle\sum^{3}_{\substack{i,j,k=1\\ i\neq j\neq k\neq i}}\displaystyle\sum^{6}_{\substack{p,q,r=4\\ p\neq q \neq r\neq p}}&\frac{W_{p+q-iipq}W_{j+k-rrjk}}{\left(\omega_{j+k-r}+\omega_r-\omega_j-\omega_k\right)}\\
&+\frac{W_{i+j-ppij}W_{q+r-kkqr}}{\left(\omega_{q+r-k}+\omega_k-\omega_q-\omega_r\right)}.
\end{array}
\end{equation}

Zakharov and Schulman discovered a parametrisation \cite{ZS82} for the six-wave resonant condition with $\omega_k \sim k^2$,
\begin{eqnarray}
k_1&=&P+R\left[u+\frac{1}{u}-\frac{1}{v}+3v\right]\nonumber,\\
k_2&=&P+R\left[u+\frac{1}{u}+\frac{1}{v}-3v\right]\nonumber,\\
k_3&=&P -\frac{2R}{u}-2Ru,\\
k_4&=&P +\frac{2R}{u}-2Ru\nonumber,\\
k_5&=&P+R\left[u-\frac{1}{u}+\frac{1}{v}+3v\right]\nonumber,\\
k_6&=&P+R\left[u-\frac{1}{u}-\frac{1}{v}-3v\right]\nonumber.
\end{eqnarray}
This parametrisation allows us to explicitly calculate $C_{123456}$ on the resonant manifold.  This is important beecause the wave kinetics take place on this manifold, that corresponds to the delta functions within the kinetic equation.
When this parametrisation is used with equation (\ref{eq:C123456}) and $W_{1234}=-\frac{1}{8}k_1k_2k_3k_4$ we find that the resonant six-wave interaction coefficient simplifies to $C_{123456}=-\frac{1}{16}k_1k_2k_3k_4k_5k_6$.  Note, that this is indeed the identical to the next term, $H_3$ in the LIA expansion with opposite sign.



\end{document}